\documentstyle[epsfig,cite]{mn}

\begin{document}

\title[Outburst of 1354$-$64]
    {The 1997 hard state outburst of the X-ray transient GS~1354$-$64 / BW Cir}
\author[Brocksopp et al.]
    {C.~Brocksopp$^1$\thanks{email: C.Brocksopp@open.ac.uk}, P.G.~Jonker$^2$, R.P.~Fender$^2$, P.J.~Groot$^{2,3}$, M.~van~der~Klis$^2$,
\newauthor
S.J.~Tingay$^4$\\
$^1$Dept. of Physics \& Astronomy, Open University, Walton Hall, Milton Keynes MK7 6AA, UK\\
$^2$Astronomical Institute ``Anton Pannekoek'' and Center for High-Energy Astrophysics, University of Amsterdam,\\
\hspace{0.5cm}Kruislaan 403, 1098 SJ Amsterdam, The Netherlands\\
$^3$Harvard-Smithsonian Center for Astrophysics, 60 Garden Street, Cambridge, MA 02138, USA\\
$^4$Australia Telescope National Facility, Paul Wild Observatory, Locked Bag 194, Narrabri 2390, NSW, Australia\\}

\date{Accepted ??. Received ??}
\pagerange{\pageref{firstpage}--\pageref{lastpage}}
\pubyear{??}
\maketitle

\begin{abstract}
We present observations of the 1997 outburst of the X-ray transient GS
1354$-$64 (BW Cir) at X-ray, optical and, for the first time, radio
wavelengths; this includes upper limits to the linear and circular
polarisation of the radio data. The X-ray outburst was unusual in that
the source remained in the low/hard X-ray state throughout; the X-ray
peak was also {\em preceded} by at least one optical outburst,
suggesting that it was an `outside-in' outburst -- similar to those
observed in dwarf novae systems, although possibly taking place on a
viscous timescale in this case. It therefore indicates that the
optical emission was {\em not} dominated by the reprocessing of X-rays
but that instead we see the instability directly. While the radio
source was too faint to detect any extended structure, spectral
analysis of the radio data and a comparison with other similar systems
suggests that mass ejections, probably in the form of a jet, took
place and that the emitted synchrotron spectrum may have extended as
far as infrared wavelengths. Finally, we compare this 1997 outburst of
GS 1354$-$64 with possible previous outbursts and also with other hard
state objects, both transient and persistent. It appears that a set of
characteristics -- such as a weak, flat spectrum radio jet, a mHz QPO
increasing in frequency, a surprisingly high optical:X-ray luminosity
ratio and the observed optical peak preceding the X-ray peak -- may be
common to all hard state X-ray transients.

\end{abstract}


\section{Introduction}

Soft X-ray transients are a class of low mass X-ray binaries in which
instabilities in the accretion disc cause a sudden increase in mass
accretion rate onto the compact object, resulting in an outburst (see
e.g. Charles 1998 and references within). While reminiscent of the
dwarf novae class of cataclysmic variables (Van Paradijs 1996) these
outbursts occur with intervals typically 10--20 years and usually
decline over several months (Tanaka \& Lewin 1995; Van Paradijs \&
McClintock 1995), sometimes displaying secondary outbursts as they do
so (e.g. the X-ray flux of GS 1124$-$684 rose again by a factor of
$\sim$ 2 about 80 days after the initial outburst. See Tanaka \&
Shibazaki 1996 and references within). In between outbursts the X-ray
transients lie in a quiescent state and this gives us an excellent
opportunity to study the companion star.

Although these objects are frequently referred to as {\em soft} X-ray
transients on account of their ultrasoft X-ray spectra during the
outbursts, an increasing number of their outbursts do {\em not} show
the soft component (e.g. GRO J1719$-$24, Van der Hooft et
al. 1996). Instead the power law component dominates the energy
spectrum, reminiscent of the low/hard X-ray state of the persistent
source Cyg X-1. However, as the term `hard X-ray transient' has
previously been used to describe the Be star+neutron star binaries
(which typically show hard X-ray spectra) we do not use this term
here. It is important to note that the term `low/hard state' is used
for historical reasons, dating back to observations of Cyg X-1 during
the 1970's -- it is somewhat misleading as its current definition (Van
der Klis 1995) is based on spectral properties and the source flux
need not necessarily be `low'.

\begin{figure*}
   \vspace*{-1cm}
   \begin{center}
   \leavevmode
   \psfig{file=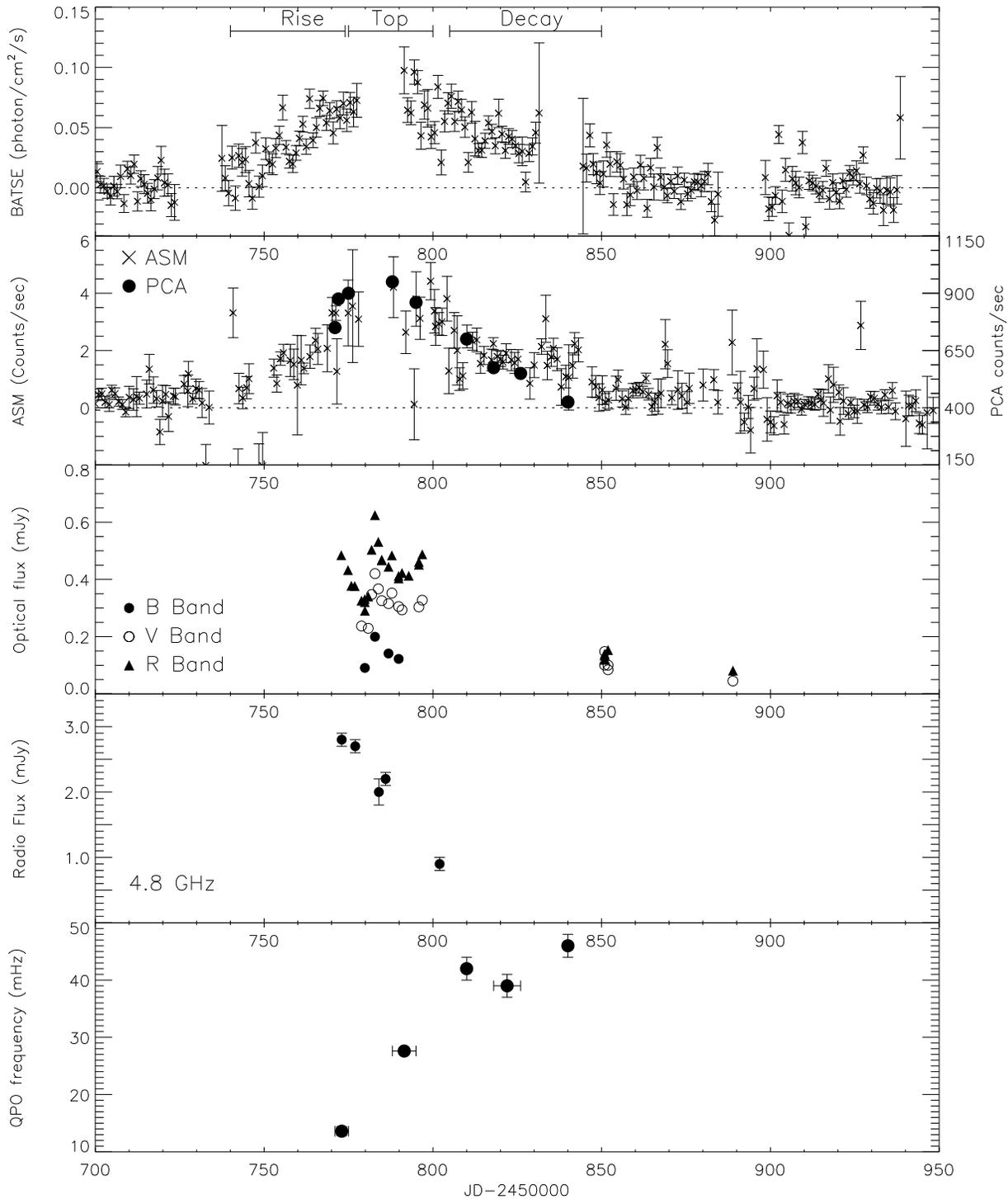, width=17cm,angle=0}
   \vspace*{-.5cm}
   \caption{BATSE, ASM, PCA, optical and radio (4.8 GHz) lightcurves of the 1997 outburst for the period 8 October 1997 -- 16 May 1998. Radio lightcurves at other frequencies are shown later in this paper. Error-bars on the optical points are smaller than the symbols. The variability of the QPO frequency is also shown in the bottom panel.}
   \label{lightcurves} 
   \end{center}
\end{figure*}

A large number of the X-ray transients have been classified as black
hole candidate (BHC) X-ray binaries -- where possible, the preferred
method is to determine the mass function (e.g. Charles 1998 and
references within) but in a large number of cases a black hole nature
has been suggested on account of the X-ray properties. Following the
unification scheme for the X-ray power density spectra of neutron
stars and BHCs proposed by Van der Klis (1994; 1995), the power
spectra of BHCs can be sorted according to the mass accretion rate;
this is thought to increase from the low/hard state, through the
intermediate state and the high/soft state, up to the very high state
(but see also Homan et al. 2000). During the low/hard state strong
band-limited noise is present. In this scheme low frequency ($\leq$ 1
Hz) quasi-periodic oscillations (QPOs) seen in the low/hard state are
seen as an aspect of the low/hard state noise. Recently, Psaltis,
Belloni, \& Van der Klis (1999) reported correlations between the
frequencies of QPOs and broad band power spectral features of neutron
star and black hole X-ray binaries.

Transitions between X-ray states have been observed during the rise
and subsequently during the decay of an outburst for a number of
sources (e.g. GS 1124$-$68 (Ebisawa et al. 1994), GRO J1655$-$40
(M\'endez et al. 1998)). Of particular interest is GX 339$-$4, in which
the transition from low/hard to high/soft X-ray states and back again
was clearly tracked at radio wavelengths, with radio emission strongly
suppressed in the high/soft state (Fender et al. 1999; Fender 2000a).

Ground based follow up observations of X-ray transients have shown
that outbursts also occur at optical and radio
frequencies. A0620$-$00, GRO J1655$-$40, GS 2023+338 (V404 Cyg) and GS
1124$-$684 are examples of X-ray transients detected in all three
wavelength regimes (Tanaka \& Lewin 1995, and references within); in
particular, the X-ray, optical and radio behaviour for GS 2023+338
were correlated during the 1989 outburst (Han \& Hjellming 1992) and
it is interesting to note that during this outburst (as in the case of
GS 1354$-$64) the source remained in the low/hard state (e.g. Tanaka
\& Lewin 1995).

\begin{table*}
\caption{Log of the $RXTE$/PCA observations. The penultimate column indicates the category of the observation (see text).} 
\label{obs_log}
\begin{tabular}{ccccccc}
\hline
Number & Observation & Date \& & Total on source & Category & Mean count\\
       &    ID       &  & observing time (ksec.) & & rate (cts/s) \\        
\hline
1 & 20431-01-01-00       & 1997-11-18 & 3.2 & rise & $\sim$750\\
2 & 20431-01-02-00       & 1997-11-19 & 6.0 & rise & $\sim$875\\
3 & 20431-01-03-00       & 1997-11-22 & 8.4 & rise & $\sim$900\\
4 & 20431-01-04-00       & 1997-12-05 & 7.6 & top & $\sim$950\\
5 & 20431-01-05-00       & 1997-12-12 & 6.4 & top & $\sim$860\\
6 & 30401-01-01-00       & 1997-12-27 & 7.0 & decay1 & $\sim$700\\
7 & 30401-01-02-00       & 1998-01-04 & 7.7 & decay2 & $\sim$575\\
8 & 30401-01-03-00       & 1998-01-12 & 6.1 & decay2 & $\sim$550\\
9 & 30401-01-04-00       & 1998-01-26 & 7.1 & decay3 & $\sim$425\\
\hline
\end{tabular}
\end{table*}

The optical emission is thought to be the result of reprocessing of
X-rays in the disc -- the disc absorbs X-rays and re-emits them at
lower energies, delaying the optical by a few seconds as reported by
Hynes (1998) in the case of GRO J1655$-$40. The radio appears to be
the result of beamed synchrotron emission taking place in relativistic
ejections of material from the centre of the disc; this may be in
response to additional material being transported through the disc by
the instability and so we might expect some correlation between the
radio and the X-ray/optical behaviour. However, a comparison between
the radio lightcurves of A0620$-$00, GS 1124$-$68 and GS 2000+25
suggests that their profiles depend on inclination angle, the number
of ejections, the time between them and their strength and speed
(Kuulkers et al. 1999). Therefore, as the radio emission may be
anisotropic in the observer's frame, unlike (presumably) X-ray and
optical, we might instead expect that it is {\em not} observed to be
so closely correlated with these two components as they are with each
other.

\subsection{GS 1354$-$64}

On 1987 February 13, the ASM on board the {\it Ginga} satellite
(Swinbanks 1987) discovered GS 1354$-$64, observing it until 1987
August (Makino et al. 1987). The X-ray data were well fit by a soft
disc blackbody component, with inner disc temperature of $\sim$0.7
keV, and a hard power law with photon index 2.1 (Kitamoto et
al. 1990). These X-ray properties are common for X-ray transients
during outburst (Tanaka and Lewin 1995) and hint at a black hole
nature for the compact object. It was also observed in the optical
during the outburst, yielding magnitudes of $V\sim16.9$, $B\sim18$
(Kitamoto et al. 1990). GS 1354$-$64 had not previously been observed
in the radio.

\begin{table*}
\caption{Identifiers, coordinates (J2000 equinox) and magnitudes of the eight reference stars used in the photometry. The `BJF' part of the identifier refers to Brocksopp, Jonker, Fender et al. 2000 (this work)}
\label{refstars}
\center
\begin{tabular}{cccc}
\hline
Identifier, RA, Dec.&$V$ (mags)&$B-V$ (mags)&$V-R$ (mags) \\
\hline
BJF J135807.07$-$644309.9&16.269 (0.005)&   0.987 (0.016)&   0.632 (0.006)\\
BJF J135807.94$-$644309.3&16.028 (0.004)&   1.111 (0.013)&   0.646 (0.005)\\
BJF J135815.53$-$644305.1&15.446 (0.002)&   1.905 (0.014)&   1.088 (0.002)\\
BJF J135814.93$-$644337.6&15.797 (0.003)&   1.570 (0.015)&   0.895 (0.003)\\
BJF J135813.21$-$644348.4&15.506 (0.002)&   1.914 (0.015)&   1.084 (0.002)\\
BJF J135809.39$-$644439.1&15.463 (0.003)&   0.889 (0.006)&   0.557 (0.003)\\
BJF J135807.71$-$644513.1&15.036 (0.002)&   0.937 (0.005)&   0.559 (0.002)\\
BJF J135811.48$-$644516.0&15.026 (0.002)&   0.811 (0.005)&   0.485 (0.002)\\
\hline
\end{tabular}
\end{table*}

The position of this source on the sky is consistent with the position
of two other transient sources; Cen X-2 (Francey et al. 1971) and MX
1353$-$64 (Markert et al. 1979). Cen X-2 was the first `soft X-ray
transient' discovered and one of the brightest (Tanaka \& Lewin
1995). The outbursts of MX 1353$-$64 (1972) and GS 1354$-$64 (1987 and
1997) reached much lower intensities and showed different X-ray
spectral properties -- if Cen X-2 and MX 1353$-$64 were the same
source as GS 1354$-$64 then it must show at least four different
states (Kitamoto et al. 1990). This is not unfeasible -- e.g. GX
339$-$4 has been observed in four states, Cyg X-1 in three (see Fender
2000a, and references within) -- it is therefore probable that these
sources are indeed one and the same, in which case we note that the
1997 outburst was significantly sub-Eddington. The error circles for
the positions of each `source' can be found in Kitamoto et
al. (1990). We investigate the nature of the various states further in
Section 4. 

GS 1354$-$64 was observed in outburst again in 1997
November. Preliminary results from the ASM on board the Rossi X-ray
Timing Explorer ({\it RXTE}) showed a rise from 16 to 50 mCrab in the
first half of November (Remillard, Marshall \& Takeshima 1997); the
BATSE instrument on the Compton Gamma Ray Observatory ({\it
CGRO}) detected hard X-rays up to 200 keV which rose from 60 to 160
mCrab (Harmon \& Robinson 1997). The optical and infrared counterparts
were also detected with magnitudes of $R\sim16.9$, $B\sim18.1$
(Nov. 22; Castro-Tirado, Ilovaisky \& Peterson 1997) and $J\sim15.35$,
$K\sim13.95$ (Nov. 20; Soria, Bessell \& Wood 1997). Optical
spectroscopy during the decay revealed strong emission lines
corresponding to H$\alpha$, H$\beta$ and He\,{\sc ii}\, $\lambda$4686
plus weaker He\,{\sc i}\,$\lambda$4471 and H$\gamma$ emission (Buxton
et al. 1998). The H$\alpha$ emission profile varied from single- to
double-peaked over three nights of observations during the decay from
outburst, indicating the presence of an accretion disc. The radio
counterpart was detected at 2.5, 4.8 and 8.6 GHz with a flux of 1--3
mJy (Nov. 24/25; Fender et al. 1997). No indication of the orbital
period was detected at any wavelength. 

A recent paper by Revnivtsev et al. (2000) gives an analysis of
$RXTE$/ASM, HEXTE and PCA data during the outburst. Their work
indicates that the X-ray energy spectrum is dominated by a power law,
typical of X-ray binaries in the low/hard state. However, the fit is
improved by the addition of a 6.4 keV iron line and a reflection
component suggestive of the presence of an accretion disc; low energy
disc photons are upscattered by a Comptonising corona to higher
energies, resulting in the X-ray emission. Here we present $B$, $V$
and $R$ photometry and radio data from the outburst, combining it with
further $RXTE$/PCA analysis and public ASM and $CGRO$/BATSE data. 

\section{Observations}

We have obtained public $RXTE$/ASM, $RXTE$/PCA and $CGRO$/BATSE data
for the 1997 outburst of GS 1354$-$64. Lightcurves for this and for
our optical and radio data are shown in Fig.~\ref{lightcurves}.

\subsection{X-ray}

The {\em RXTE} satellite observed GS 1354$-$64 ten times with the
onboard proportional counter array (PCA; Jahoda et al. 1996). We
present only the analysis of the first nine of these since the count
rates were too low and the observing time too short during the last
observation to constrain our fit parameters. During the outburst a
total of nearly 60 ksec of good data was obtained. A log of the
observations we present here (following background corrections) can be
found in Table~\ref{obs_log}. During $\sim$4\% of the time only 4 of
the 5 PCA detectors were active -- when this was the case the data
were averaged. The count rates quoted are approximate because they
depend on the background model and can vary over an observation. 

All the data were obtained in each of three modes. The Standard 1 mode
has a time resolution of 1/8 s in one energy channel (2--60 keV). The
Standard 2 mode has a time resolution of 16 s and the effective 2--60
keV PCA energy range is covered by 129 energy channels. In addition,
high time resolution data (with a time resolution of 125 $\mu$s or
better) were obtained for all observations in at least 64 energy
channels covering the effective 2--60 keV PCA energy range.  

Additional soft (2--12 keV) X-ray data has been obtained from the ASM
on board {\it RXTE}; we use the public archive data from the web
(http://xte.mit.edu). A detailed description of the ASM, including
calibration and reduction is published in Levine et al. (1996). Our
hard  (20--100 keV) X-ray data came from the BATSE instrument on the
Compton Gamma Ray Observatory ({\it CGRO}) and was processed using the
standard BATSE earth occultation software. Again, we have used the
public archive data from the web. A detailed overview of the BATSE
instrument can be found in Fishman et al. (1989). 

\subsection{Optical}

The optical counterpart to GS 1354$-$64 was observed on 18 nights
during 1997 November/December and 2 nights the following
February. Further observations taken a year later revealed no source
suggesting that the source in quiescence has $V\ge$22 magnitudes
($\le$ 0.005 mJy). The 0.91m Dutch telescope at the European Southern
Observatory in Chile was used, equipped with a 512$\times$512 TEK CCD
and standard Johnson $B$ and $V$, and Cousins $R$ filters. Typical
exposure times were 20 minutes in B and 5 minutes in V and R (although
there were also some 10 minute observations in all three bands).  

The data were processed using routine bias subtraction and flat field
division in {\sc iraf}. Aperture photometry was applied, both to GS
1354$-$64 and to eight field stars; the coordinates of these stars are
shown in Table~\ref{refstars}. The photometry was then calibrated
(also with {\sc iraf}) using photometric standard stars in the fields
of PG 0231+051 and Mark A (Landolt 1992). Magnitudes are given in
Table~\ref{phot}. The magnitudes were converted into mJy to ease
comparison with the radio fluxes; Johnson conversions were used for
the $B$ and $V$ bands, Cousins for the $R$ band.  

\begin{table}
\caption{Calibrated magnitudes for our optical photometry. Errors are given in parentheses.}
\label{phot}
\begin{tabular}{cccc}
\hline
JD$-$2450000&$B$&$V$&$R$\\
\hline
772.87    &   --   &    --     &    16.89 (0.04)\\
774.86   &  --     &  --       &    17.02 (0.02)\\
775.86   &    --    &   --     &    17.16 (0.01)\\
776.86   &      -- &   --     &     17.17 (0.02)\\
778.86   & --      &  18.02 (0.07)& 17.32 (0.02)\\
779.84  &    --    &   --        &  17.34 (0.03)\\
779.85  &      --   &   --       &  17.30 (0.01)\\
779.86& 19.18 (0.07)&   --    &     17.45 (0.06)\\
780.85&  --         & 18.06 (0.02)& 17.28 (0.01)\\
781.85&    --       & 17.61 (0.03)& 16.85 (0.01)\\
782.84& 18.32 (0.02)& 17.40 (0.03)& 16.62 (0.01)\\
783.84&  --        &  17.55 (0.01)& 16.79 (0.01)\\
784.85&    --      &  17.68 (0.01)& 16.93 (0.01)\\
784.86&      --     &    --      &  16.93 (0.01)\\
786.86& 18.70 (0.03)& 17.71 (0.07)& 16.99 (0.02)\\
787.85&  --        &  17.59 (0.01)& 16.89 (0.01)\\
789.85&    --        &     --    &  17.07 (0.02)\\
789.85& 18.85 (0.14)& 17.75 (0.02)& 17.09 (0.01)\\
790.84&  --       &   17.79 (0.01)& 17.04 (0.01)\\
792.85&    --     &       --      & 17.07 (0.05)\\
795.84&--         &   17.75 (0.02) &16.95 (0.03)\\
795.84&  --       &         --    & 16.97 (0.01)\\
796.84&    --     &   17.67 (0.02)& 16.89 (0.01)\\
850.88&--        &       --       & 18.41 (0.22)\\
850.88&  --     &     18.95 (0.40)& 18.28 (0.10)\\
850.88&    --   &        --       & 18.29 (0.06)\\
850.88&--       &     18.53 (0.31)& 18.24 (0.14)\\
851.89&  --     &     18.95 (0.02)& 18.14 (0.01)\\
888.88&    --   &     19.82 (0.14)& 18.83 (0.06)\\
\hline
\end{tabular}
\end{table}

\begin{table*}
\caption{ATCA observing log, and results of image-plane (naturally weighted) and {\it uv}-plane point-source fits to the data.}
\label{radio-data}
\center
\begin{tabular}{ccccccc}
\hline
Date          & Frequency & Total     & Image plane fit & {\it uv}
plane fit&Linear Pol.&Circular Pol/.\\
(JD$-$2450000)&(GHz)      & Time (hrs)& (mJy) & (mJy)& ($3\sigma$)& ($3\sigma$)\\
\hline
773      & 1.384 & 2.5 & $2.6 \pm 0.2$ & $3.6 \pm 0.1$&$<$8\%&$<$7\%\\
         & 2.496 &     & $2.8 \pm 0.1$ & $2.9 \pm 0.1$&$<$9\%&$<$8\%\\
         & 4.800 &     & $2.8 \pm 0.1$ & $2.9 \pm 0.1$&$<$6\%&$<$6\%\\
         & 8.640 &     & $1.6 \pm 0.1$ & $1.9 \pm 0.1$&$<$15\%&$<$16\%\\
\hline
777--778 & 1.384 & 0.7 & $2.5 \pm 0.5$ & $2.3 \pm 0.3$&&\\
         & 2.496 &     & $2.5 \pm 0.1$ & $3.7 \pm 0.2$&&\\
	 & 4.800 &     & $2.7 \pm 0.1$ & $3.7 \pm 0.3$&&\\
	 & 8.640 &     & $2.5 \pm 0.1$ & $2.2 \pm 0.2$&&\\
\hline
784--785 & 2.368 & 0.58& $3.5 \pm 0.1$ & $3.7 \pm 0.3$&&\\
	 & 4.800 &     & $2.0 \pm 0.2$ & $2.7 \pm 0.2$&&\\
	 & 8.640 &     & $1.5 \pm 0.1$ & $2.2 \pm 0.3$&&\\
\hline
786--787 & 4.800 & 1.25& $2.2 \pm 0.1$ & $3.1 \pm 0.2$&& \\
         & 8.640 &     & $1.8 \pm 0.2$ & $3.9 \pm 0.5$&& \\
\hline
802      & 4.800 & 0.75& $0.9 \pm 0.1$ & $1.7 \pm 0.4$&&\\
         & 8.640 &     & $1.1 \pm 0.1$ & $1.8 \pm 0.4$&&\\
\hline
\end{tabular}
\end{table*}

\subsection{Radio}

Radio observations were made at the Australia Telescope Compact Array
at 1.384, 2.496, 4.800 and 8.640 GHz at eight epochs during the period
JD 2450773 -- JD 2450802 (1997 November 20, 24, 25 and December 1, 2,
3, 4, 19). The array was in configuration 6C, which included the 6 km
antenna, giving a nominal angular resolution of $\sim 1$ arcsec at
8.640 GHz.  Standard flagging, calibration and imaging techniques were
used within {\sc miriad}; the flux calibrator was PKS 1934$-$638 and
the phase calibrator was PMN J1417$-$5950. The integration times for the
first epoch totalled 2.5 hours at each frequency; typical integration
times for following epochs were $<$1 hour, yielding reasonable S/N but
unfortunately poor $uv$ coverage (the {\it uv} plane is the projection
of the antenna baselines onto the plane of the sky). As a result, for
most epochs the synthesised beam from the observations was extremely
elongated. Furthermore the sparse {\it uv} coverage meant that the
flux densities measured were a function of the cell size used in
making the maps, as a result of different gridding in the FFTs. Both
effects resulted in difficulty in measuring reliably the flux
densities at the epochs with the shortest observations. In order to
improve the situation, we combined data from two adjacent days (for
Nov 24/25, Dec 01/02 and Dec 03/04), which significantly improved the
{\it uv} coverage in all three cases. In addition, as well as
performing point-source fits to the final (naturally weighted) images,
we also performed point-source fits in the {\it uv} plane, thereby
avoiding gridding errors. On Dec 3/4 data were collected in two
adjacent bands, centred on 2.240 and 2.496 GHz respectively, and
combined to improve the S/N by $\sqrt2$ at an effective frequency of
2.368 GHz. 

In Table~\ref{radio-data} we present the ATCA observing log and
results of our image-plane and {\it uv}-plane fits to a point
source. It is clear that both the image-plane and {\it uv}-plane
results follow a similar trend, steadily declining by a factor $\sim
2$ in the month between Nov 20 and Dec 19. It is also clear that the
spectrum is generally flat, especially at lower ($\leq 4.800$ GHz)
frequencies. Between 4.800 -- 8.640 GHz the spectrum appears to be
initially optically thin, but gradually flattens until becoming
inverted by the last detection on Dec 19. We note that the {\it
uv}-plane fits invert one epoch earlier than those of the image plane.
 

\section{Results and Analysis}

\subsection{X-ray}

The X-ray lightcurves in Fig.~\ref{lightcurves} have a triangular
shape, as defined by Chen, Shrader \& Livio (1997) with a possible
secondary outburst during the decay at $\sim$ JD 2450835. It is clear
from Fig.~\ref{lightcurves} that the soft and hard X-rays are well
correlated -- we obtain a value of 0.6 for the Spearman rank
correlation coefficient between the daily averaged {\it RXTE}/ASM and
{\it CGRO}/BATSE lightcurves, despite the poor S/N. We also note that
the energy spectra of both X-ray bands are dominated by a power law
component (Harmon \& Robinson 1997, Revnivtsev et al. 2000) which
indicates a common origin, i.e. a Comptonising corona which upscatters
low energy (optical and uv) disc photons to higher energies (e.g. Van
Paradijs 1998). This is significantly different from the 1987
outburst, when the X-rays showed a soft disc blackbody spectrum
(Kitamoto et al. 1990). In the 1997 outburst the presence of an
accretion disc can be inferred only indirectly with the inclusion of a
reflection component in the soft X-ray spectrum (Revnivtsev et
al. 2000).

\begin{figure}
   \begin{center}
   \leavevmode
   \psfig{file=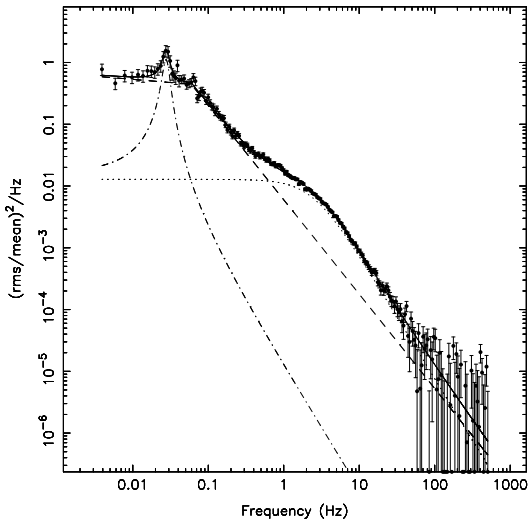,width=8cm}
   \vspace{0.5cm}
   \caption{Normalized (Belloni \& Hasinger 1990) power spectrum of
   observations 4 and 5 combined. The power arising in the power
   spectrum due to Poisson noise has been subtracted. The solid line
   represents the best fit to the data.This best fit function is built
   up by three components. The dashed line represents the contribution
   of the broken power law component, the dotted line represents the
   contribution of the Lorentzian component with its frequency fixed
   at 0 Hz, and the dashed-dotted line represents the contribution of
   the Lorentzian used to represent the QPO.}
   \label{top_qpo} 
   \end{center}
\end{figure}

For all {\it RXTE}/PCA observations we calculated power spectra with a
Nyquist frequency of 512 Hz using data segments of 512 s length each
in one combined broad energy band ranging from 2--60 keV. To
characterise the power spectra (1/512--512 Hz) we used a fit function
consisting of a broken power law, plus a Lorentzian with its frequency
fixed at 0 Hz to describe the low frequency noise component. We also
included a second Lorentzian to describe the QPO on top of the noise
component. The frequency of the noise Lorentzian became negative when
it was treated as a free parameter. Errors on the fit parameters were
determined using $\Delta\chi^2$=1.0 ($1 \sigma$ single
parameter). Power arising in the power spectrum due to Poisson noise
has been subtracted -- this also takes deadtime into account. The fit
to the normalised (Belloni \& Hasinger 1990) power spectrum of
observations 4 and 5 combined is shown in Fig.~\ref{top_qpo}.

We divided the outburst into three parts on the basis of the X-ray
lightcurve: the rise, the top, and the decay (see Table~\ref{obs_log},
Fig.~\ref{lightcurves}). The decay observations were further
subdivided into three parts in order to follow the changes in the fit
parameters during the decay of the outburst.  

\begin{table*}
\caption{Properties of the two Lorentzians and the broken power law
component. The frequency of the noise component was fixed at 0 Hz.} 
\label{qpo_tab} 
\begin{tabular}{lccccc}
\hline
Category & Rise & Top & Decay1 & Decay2 & Decay3 \\
\hline
rms QPO (\%) & $10\pm2$ & $10\pm1$ & $ 10\pm2$ & $10\pm2 $ & $9\pm2$\\
FWHM QPO (mHz) & $4.4\pm1.5$ & $7\pm1$ & $11^{+11}_{-6}$ & $19\pm7$ &$11^{+10}_{-4}$\\
$\nu_{QPO}$ (mHz) & $13.6\pm0.8$ & $27.6\pm0.6$ & $42\pm2$ & $39\pm2$& $46\pm2$\\
rms noise (\%) & $23.1\pm0.2$ & $22.1\pm0.2$ & $21.8\pm0.5$ &$24.1\pm0.3$ & $25.7\pm0.5$\\
FWHM noise (Hz) & $3.84\pm0.06$ & $4.87\pm0.07$ & $5.1\pm0.1$ &$4.8\pm0.1$ & $4.6\pm0.2$\\
rms break (\%)& $33.4\pm0.6$ & $27.7\pm0.4$ & $27.3\pm0.7$ &$26.3\pm0.4$ & $26.8\pm0.8$\\
$\nu_{break}$ (mHz) & $56\pm5$ & $60\pm4$ &  $91\pm7$ & $96\pm4$ &$110\pm10$\\
$\alpha$ ($\nu<\nu_{break}$) & $0.59\pm0.07$ & $0.100\pm0.001$ & $0.12\pm0.05$ & $(-3\pm5)\times10^{-2}$ & $0.22\pm0.07$\\
$\alpha$ ($\nu>\nu_{break}$) & $1.46\pm0.01$ &  $1.53\pm0.03$ & $1.55\pm0.05$ & $1.82\pm0.05$ & $1.9\pm0.1$\\
\hline
\end{tabular}
\end{table*}

The properties of our fit parameters are given in
Table~\ref{qpo_tab}. We detect a QPO at low frequencies; its
frequency increased from $13.6\pm0.8$ mHz at the rise to $46\pm2$ mHz
at the end of the decay. The FWHM also increased over this range from
$4.4\pm1.5$ mHz to $11^{+10}_{-4}$ mHz. The break frequency increased
gradually from $56\pm5$ mHz during the rise to $110\pm10$ mHz during
the last part of the decay. The power law index at frequencies below
the break frequency changed from $0.59\pm0.07$ during the rise to
$0.100\pm0.001$ at the top, during the decay it varied between
$0.22\pm0.07$ and $(-3\pm5)\times10^{-2}$.  At frequencies above the
break frequency the power law became gradually steeper with indices of
$1.46\pm0.01$ during the rise to $1.9\pm0.1$ during the last part of
the decay.

The increasing of the frequency at which the QPO and break occur
suggest that the inner radius of the disc is decreasing as the
outburst takes place (Revnivtsev et al. 2000). A general softening of
the spectrum, consistent with this geometry, was also observed and was
seen previously in GS 1354$-$64 by Kitamoto et al. (1990). A similar
increase in frequency of low-frequency QPOs was discovered during an
outburst of the X-ray transient GRO J1719$-$24 by Van der Hooft et
al. (1996). If the increasing frequencies of the QPOs really do relate
to the inner disc then it is interesting to note that the inner disc
radius does not appear to increase again during the decay from
outburst. 

From the high fractional rms amplitude of the low frequency noise
throughout the outburst, the hard spectrum (Revnivtsev et al. 2000)
and the relatively low intensity increase (compared with the outbursts
of Cen X-2 (Francey et al. 1971) and MX 1353$-$64 (Markert et
al. 1979) --  we investigate this further in Section 4) we conclude
that the source was in the low/hard state during the entire outburst
(see for reference Van der Klis 1995). This is unusual as the majority
of X-ray transients enter the high/soft state during outbursts; known
exceptions include GRO J0422+32 (e.g. Sunyaev et al. 1993), GRO
J1719$-$24 (e.g. Van der Hooft et al. 1996) and GS 2023+338, although
in the latter case it is possible that the system was in a soft but
absorbed state at its peak, then returning to the low/hard state
(\.Zycki, Done \& Smith 1999).  

While the increase in QPO frequency and in break frequency during the
decay of the outburst is {\em not} a common feature of BHCs whilst in
the low/hard state, it appears that it {\em is} a typical feature
during the {\em outbursts} of X-ray transients (see Section
4). Observations of the persistent BHCs during {\em transitions} from
the low/hard to the high/soft state would be useful to determine
whether or not the QPO frequency increases at these times -- {\it
RXTE} observations of the 1996 Cyg X-1 transition to the soft
(intermediate?) state suggest that although a mHz QPO was present and
the frequency certainly varied, it was not simply a frequency increase
during the transition (Cui et al. 1997).

\subsection{Optical}

Our optical observations show that the X-ray increase was accompanied
by an optical outburst reaching $B\sim18.3$, $V\sim17.3$ and
$R\sim16.6$ magnitudes at maximum. As is typical for X-ray transients
(e.g. Van Paradijs \& McClintock (1995) and references within), the
subsequent decline was considerably longer than that of the X-rays,
taking at least 120 days. Further observations a year later revealed
no source, suggesting that there had been a brightening of $\ge$ 5
mags. above the quiescent level. 

\begin{figure}
   \begin{center}
   \leavevmode
   \hspace*{-1cm}\psfig{file=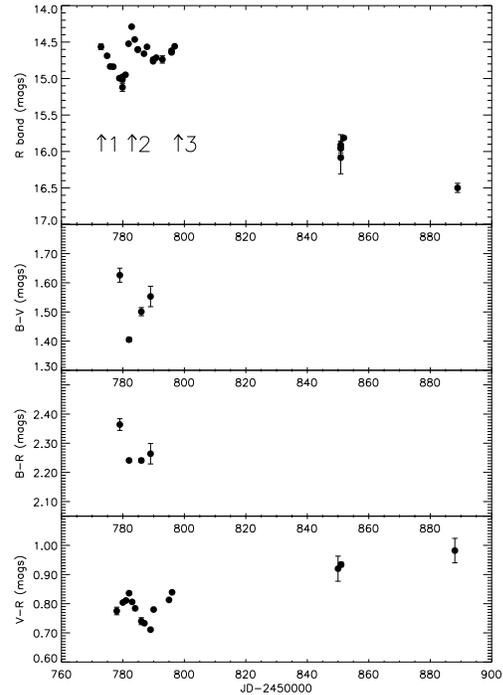,width=6.5cm, angle=0}
   \vspace{0.5cm}
   \caption{Dereddened $R$-band photometry (corrected for interstellar extinction using $E(B-V)\sim1$ (Kitamoto et al. 1990) and the $A_{\lambda}/A_V$ relations of Cardelli, Clayton \& Mathis (1989)) and $B-V$, $B-R$ and $V-R$ colour evolution (instrumental magnitudes were used to determine colours in order to minimise errors). Arrows on the $R$-band plot indicate the three apparent peaks in the photometry.}
   \label{colours-evolution} 
   \end{center}
\end{figure}

The photometry suggests that the optical and X-ray events were not as
well correlated as would be expected in a `normal' soft X-ray
transient event, in which the dominant source of optical emission
would be reprocessing of X-rays. Indeed, whereas the X-rays produce a
gradual rise and decay the optical photometry appears to peak at least
three times. The three apparent peaks are labelled on
Fig.~\ref{colours-evolution}.

The first `peak' occurs prior to our observations, but can be inferred
from the $R$-band decline at the beginning of the dataset. The X-rays
have begun to rise by this point but there is still a $\ge$10 day
delay after our initial $R$ band observations before the X-ray peak is
reached. This optical peak is unlikely to be the result of
reprocessing of soft disc X-rays on account of the low X-ray flux and
the lack of evidence for a disc component in the X-ray energy
spectrum. (It is not impossible, despite the lack of correlation
e.g. GRO J1655$-$40 shows an anti-correlated X-ray/optical outburst
but echo mapping shows that reprocessing of X-rays is still the
dominant contributor to the optical flux (Hynes et al. 1998).)

A similar phenomenon is seen in dwarf novae -- as the disc instability
moves inwards from the outer parts of the disc, the outburst is seen
first in the optical and then in the ultraviolet. While it should be
feasible that this phenomenon occurs also in X-ray transient events,
it has rarely been observed (e.g GRO J1655$-$40, Orosz 1997). If this
is the case for this outburst of GS 1354$-$64 then our data suggest
that the instability takes $\geq$10 days to cross the disc and peak in
the X-rays. This is not impossible -- Orosz (1997) calculates an
optical lead of 6 days -- but would place fairly large lower limits on
the size of the orbit and/or donor star and would also suggest that
the instability moves on a viscous timescale, unlike the thermal
instabilities of the dwarf novae. The outburst of GRO J1655$-$40 was
different from that of GS 1354$-$64 in that it reached the high/soft
state (possibly the very high state (M\'endez et al. 1998)), it was
considerably more luminous at all wavelengths and the profiles of its
rise to maximum in the BATSE and ASM data for GRO J1655$-$40 were
considerably longer. In particular, the hard X-ray rise occurred
$\sim30$ days after the soft X-rays -- a phenomenon not seen in the
case of GS 1354$-$64, where the ASM and BATSE data appear well
correlated on account of both X-ray bands being dominated by the power
law component. GRO J1655$-$40 may therefore not be a suitable system
with which to compare GS 1354$-$64.

The second peak (and only peak for which both the rise and decay were
observed) took place during the `top' period of the X-ray lightcurve
($\sim$JD2450784) and was observed in all three optical bands. Without
full X-ray coverage and better S/N in the X-ray bands it is not
possible to say whether there is any evidence for correlated
X-ray/optical behaviour. There is the hint of a small decline in the
X-rays coincidentally with the decay of this second optical peak but
the S/N does not allow us to be conclusive. If the optical and X-ray
were correlated then we might assume that the second optical peak was
produced by the reprocessing of X-rays (although the hard X-ray energy
spectrum suggests that this is unlikely); a non-correlation would
suggest that a second instability had passed through the
optical-emitting regions of the disc.

The $V$ and $R$ bands also suggest the presence of a third rise in the
lightcurve, $\sim$15 days after the time of the first maximum; it may
also be hinted at in the X-rays, although a more significant secondary
X-ray outburst takes place during the decay around JD 2450835. These
small secondary maxima during the decay of an outburst are seen
commonly in X-ray transients (e.g. GRO J0422+32, Callanan et al. 1995)
and may be due to additional, smaller instabilities.

Using the reddening correction determined by Kitamoto et al. (1990),
$E(B-V)\sim1$, and the $A_{\lambda}/A_V$ relations of Cardelli,
Clayton \& Mathis (1989) we have corrected our photometry for
interstellar extinction ($A_B\sim4.1$, $A_V\sim3.1$, $A_R\sim2.3$). In
Fig. ~\ref{colours-evolution} we plot dereddened $R$-band magnitudes
and the intrinsic colour evolution over time. To minimise errors the
colours have been determined from our original instrumental
differential magnitudes.

We have no colour information for peak 1 but the $B-V$, $B-R$ and
$V-R$ colours for peak 2 all hint at being anti-correlated with the
photometric lightcurve, although there may be a delay of $\sim0.5$
days in the case of the latter. (Alternatively it is possible that the
$V-R$ colours are actually anti-correlated with the other colours, in
which case there must be an additional contribution to $V-R$ from some
other component, perhaps the jet.) The colours also suggest that the
source became bluer during the rise to peak 2; this could be explained
by a disc instability moving inwards as the temperature and density of
the disc will be increasing inwards. On the decay from peak 2 the
colours redden, consistent with the disc cooling. We notice that the
$B-V$ and (to a lesser extent) the $B-R$ colours vary more
significantly than $V-R$; this suggests that, while the whole disc is
brightening, the hot inner regions do so in particular and this causes
a greater brightening in the $B$ band than the other two. This would
be expected and shows that at shorter wavelengths the disc spectrum
dominates the optical emission. Clearly improved coverage of future
outbursts would be extremely beneficial in order to confirm these
possibilities.

Although we have only one pair of B band/ASM points (and these are
separated by 1.5 days), we also calculate the ratio of X-ray to
optical luminosity, $\xi=B_0+2.5\log F_x$($\mu$Jy), where $B_0$ is the
extinction corrected $B$ magnitude and $F_x$ is the X-ray flux (Van
Paradijs \& McClintock 1995). Assuming $E(B-V)\sim1$, an interstellar
extinction ($A_B$) of 4.1 mags. and 1 crab $\sim$ 75 counts/sec
($RXTE$/ASM) $\sim$ 1060 $\mu$Jy we obtain a value of $\xi=19$. This
is slightly lower than that of other X-ray transients, indicating that
the source might be under-luminous in the X-rays -- typically,
$\xi=22$ ($\pm1$). We note that during the 1987 outburst the ratio was
calculated to be $\sim 16$. This is surprisingly low and was thought
likely to be due to errors in the reddening corrections and/or
non-simultaneity of the observations (Van Paradijs \& McClintock
1995). During the 1997 outburst it appears that the optical emission
was dominated by the viscous instability moving through the disc, with
minimal reprocessing. Other potential sources for the optical excess
are synchrotron emission from the jet and/or a possible contribution
due to irradiation of the companion star, although this may be
insignificant given the low magnitude of the source in quiescence.

\subsection{Radio}

\begin{figure}
   \begin{center}
   \leavevmode
   \psfig{file=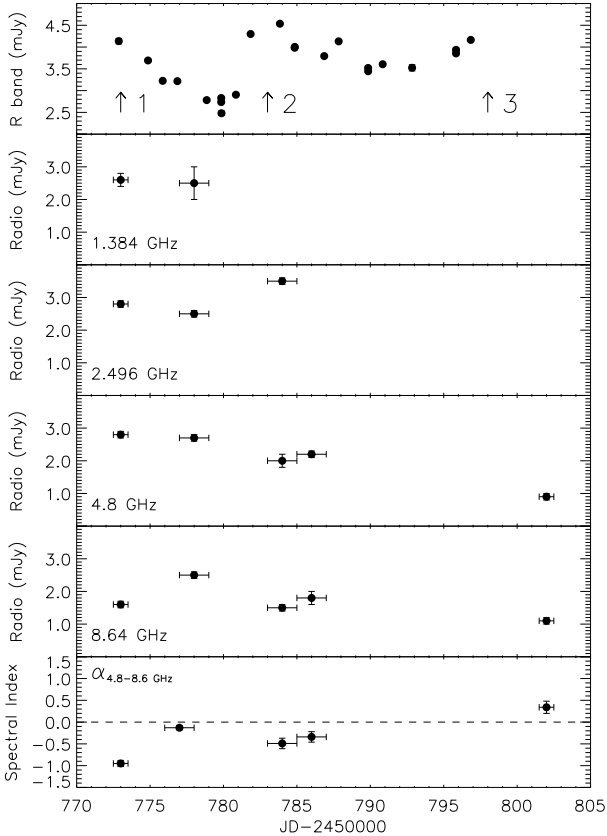, width=8cm}
   \vspace{0.5cm}
   \caption{The $R$-band photometry and radio lightcurves at 1.4, 2.5, 4.8 and 8.6 GHz on an expanded scale. Evolution of the spectral index ($\alpha$) at 4.8 and 8.6 GHz is shown in the bottom panel. Arrows on the $R$-band plot indicate the three apparent peaks in the optical photometry.}
   \label{radio} 
   \end{center}
\end{figure}

We have observed the radio counterpart to GS 1354$-$644 for the first
time. Detected at 1.384, 2.496, 4.8 and 8.64 GHz it is a weak source
of 1--3 mJy and our radio maps show no evidence for extension --
combining data from the first two epochs suggests that the upper
limits to the source extension are 11.2, 6.9 and 4.9 $\times10^3$ AU
at 2.5, 4.8 and 8.6 GHz respectively, assuming a distance of $\sim10$
kpc (Kitamoto et al. 1990). This is not really surprising given such a
low flux and the rather short exposure times of our observations. We
have also determined the radio position of GS 1354$-$644 by assuming a
point source and fitting a Gaussian -- by assuming an error of 200 mas
on the position of the phase calibrator we calculate the position of
GS 1354$-$644 to be R.A.: 13:58:09.7, dec: $-$64:44:05.8 ($\pm200$
mas) at 4.8 GHz -- the positional error is dominated by the
uncertainty in the absolute coordinates of the phase calibrator.

Lightcurves at all four observing frequencies are shown in
Fig.~\ref{radio}. Surprisingly, the radio frequencies are not well
correlated with each other -- although they all seem to follow similar
trends, in that some rise has taken place prior to our observations
and a second rise may have occured just after the second optical
peak. Exact time differences cannot be measured as each radio point
contains data spanning up to two days. The lack of correlation with
the X-rays and optical photometry is perhaps surprising and possibly
due either to our lack of data coverage or to reasons given by
Kuulkers et al. (1999). Each maximum is probably associated with an
ejection event as seen in e.g. A0620$-$00 (Kuulkers et al. 1999).

While we would expect a certain frequency dependency in the relative
delays and luminosities for these ejection events, the lack of obvious
correlation is surprising. Correlated radio/optical behaviour has been
observed previously (e.g. GS 2023+338, Han \& Hjellming 1992) and it is
possible that had we better radio coverage and/or S/N then the same
might be true here for all radio frequencies.

Fig.~\ref{radio} also shows the evolution of the spectral index
($S_{\nu} \propto \nu^{\alpha}$ where $\alpha$ is the spectral index,
$\nu$ is the frequency and $S_{\nu}$ is the flux density at $\nu$ in
mJy) over the period of our observations. The spectral indices are
also tabulated in Table~\ref{alpha} and were determined by fitting the
above power law to the data at 4.8 and 8.6 GHz -- the other
frequencies were not used to calculate the spectral index as they were
not measured at every epoch. Clearly $\alpha$ is generally negative
particularly at the first epoch, corresponding to optically thin
synchrotron emission. The spectrum becomes inverted at the final epoch
which is probably consistent with a partially self-absorbed continuous
jet as seen in Cyg X-1 and GX 339$-$4 -- it seems that this spectrum
inversion is a typical feature of the jets of low/hard state systems
(e.g. Fender 2000a and references within). Alternatively the spectrum
inversion could be indicative of a plasmon ejection emitting optically
thick synchrotron emission which will evolve to optically thin at
successively lower frequencies as the plasmon expands (see
e.g. Kuulkers et al. 1999 for further evidence of this phenomenon) --
however this is more typically seen in {\em soft} X-ray transient
outbursts. We note that using the $uv$ plane fluxes from
Table~\ref{radio-data} results in the spectrum inverting one epoch
earlier.

\begin{table}
\center
\caption{The spectral index calculated for each epoch of our radio observations. Only the 4.8 and 8.6 GHz data have been used so as to be consistent at every epoch.}
\label{alpha}
\begin{tabular}{ccc}
\hline
JD$-$2450000&$\alpha$&error\\
\hline
773& $-$0.95&  0.07\\
777/8& $-$0.13&  0.05\\
784/5& $-$0.49&  0.12\\
786/7& $-$0.34&  0.12\\
802& 0.13&     0.14\\
\hline
\end{tabular}
\end{table}

By combining data from adjacent epochs, we have improved the S/N
sufficiently to measure upper limits to the linear and circular
polarisation. These limits are determined from fits to the image plane
with percentages calculated as a fraction of the (image plane) total
flux density; they can be found in Table~\ref{radio-data}. It is
interesting to note that linear polarisation of the soft X-ray
transient 4U 1630$-$47 (Flux density $\sim$ 0.5--3 mJy) was detected at
27\% i.e. a level that we would have picked up had it been the case
for GS 1354$-$64. However, observations of GS 2023+338 (Han \&
Hjellming 1992) and GX 339$-$4 (Corbel et al. 2000) in the {\em hard}
state yielded linear polarisations of 1-4\% and this may turn out to
be a value common to the hard state sources (which are probably
partially self-absorbed).

\begin{figure}
   \begin{center}
   \leavevmode
   \hspace{-0.75cm}\psfig{file=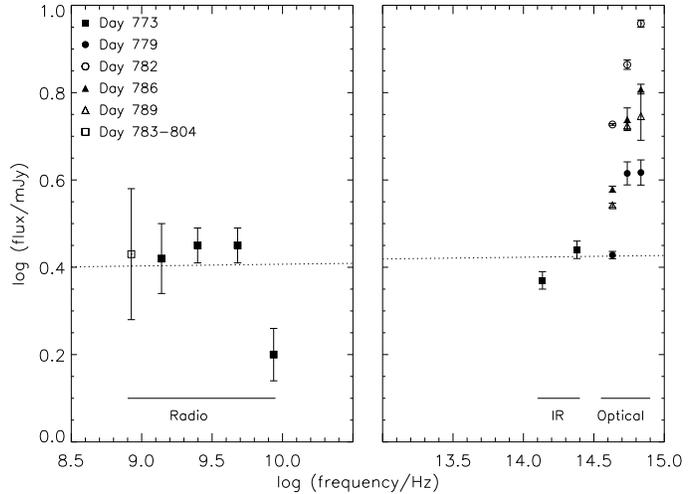, width=6.5cm,angle=90}
   \vspace{0.5cm}
   \caption{Spectrum ranging from radio through to optical. We include the infrared ($J$ and $K$ band) points of Soria, Bessell \& Wood (1997) and the MOST data of Hunstead \& Campbell-Wilson (private communication). Optical and infrared data have been corrected for interstellar extinction using a reddening estimate of $E(B-V)=1$ (Kitamoto et al. 1990) and values of $A_{\lambda}/A_V$ from Cardelli, Clayton \& Mathis (1989). The dotted line is the best fit straight line through the radio, infrared and $R$-band ($\alpha\sim0.004$)}
   \label{spectrum} 
   \end{center}
\end{figure}

While our optical investigation has suggested that the optical
emission is the result of one or more heating waves travelling inwards
through the disc, it is also interesting to note that the optical
fluxes are comparable with those of the radio, suggesting that the
flat spectrum associated with the central radio source might extend to
higher frequencies. A flat synchrotron spectrum beyond the radio
regime has been seen in a number of X-ray binaries in the low/hard
X-ray state e.g. Cyg X-1 to the mm wavelengths and GRS 1915+105 to the
infrared (Fender et al. 2000). It is therefore feasible that there may
be a contribution to the optical emission from the jet; with the
possible correlation between the optical and the radio this would not
be surprising. 

\begin{table*}
\center
\caption{Comparison between outbursts of Cen X-2, MX 1353$-$64 and GS 1354$-$64, which may all be the same source. X-ray peaks are quoted for the 1-10 keV range. See Tanaka \& Lewin (1995) and references within.}
\label{cenx2}
\begin{tabular}{ccccccc}
\hline
Source&Date&Duration&Soft X-ray Peak&X-ray State& Optical Peak\\
&&(months)&(mCrab)&&(mags)\\
\hline
Cen X-2&1966--1967&$\sim8$&$\sim8000$&(Very?) high/soft&$V<13.5$ (WX Cen)?\\
MX 1353$-$64&1972&$\sim11$&75&low/hard&--\\
GS 1354$-$64&1987&$\sim6$&$\ge300$&high/soft&$V\le16.9$ (BW Cir)\\
GS 1354$-$64&1997&$\sim5$&50&low/hard&$V\le17.3$ (BW Cir)\\
\hline
\end{tabular}
\end{table*}

%

To investigate this further we plot the extinction corrected spectrum
for our optical data, plus the published infrared ($J$ and $K$ band)
points of Soria, Bessell \& Wood (1997) and the radio fluxes quoted in
Table~\ref{radio-data} which are from the same epoch as the IR
(Fig.~\ref{spectrum}). We have also extended the spectrum down to 843
MHz with the inclusion of a data point from the Molonglo Observatory
Synthesis Telescope (MOST) (Hunstead \& Campbell-Wilson, private
communication). This plot indicates that while the $B$ and $V$ band
data are too bright to be associated with the radio, the $R$ band and
the infrared points seem to lie approximately in a straight line --
the best fit spectral index to the radio, infrared and $R$ band points
is $0.004\pm0.01$, which is consistent with flat synchrotron emission
up to the near infrared. In contrast to this, the optical data alone
yields a spectral index of 1.12--1.68, depending on the epoch, and the
$B$ and $V$ data alone yields $\alpha\sim $0.02--0.97. With a
theoretical thermal disc spectrum (with no irradiation) of
$S_{\nu}\propto\nu^{1/3}$ (e.g. Frank, King, Raine, 1992 and
references within) it is probable that there may be some additional
optical component causing the steeper optical spectrum at some
epochs. However, we note that without full coverage of the spectrum
through the sub-mm and infrared regimes we cannot be certain the
spectrum really is flat from radio to near-infrared. 

We note that the the 8.64 GHz point is somewhat discrepant. While it
is possible that there is some systematic error due to the poor {\it
uv} coverage, we point out that it may take some time after the
ejection for the spectrum to invert, i.e. become dominated by a steady
self-absorbed flow. For example the radio spectrum of V404 Cyg did not
invert until $\sim$ 25 days after the initial (hard state) outburst
(Fender 2000b, Han \& Hjellming 1992).

\section{Comparison with other outbursts}
\subsection{Previous (possible) outbursts of GS 1354$-$64}

We summarise the four possible outbursts of GS 1354$-$64 in
Table~\ref{cenx2}. It is clear that the characteristics (i.e. the
spectrum and luminosity) of each outburst are very different --
Kitamoto (1990) concluded that for GS 1354$-$64 to have produced all
three outbursts known by that date, then the source must show four
distinct X-ray states, i.e. very high, high/soft, low/hard and
faint/off states. Although less bright, the 1997 outburst appears to
show a similar state to that of the 1972 event reported by Markert et
al. (1977). However, as Kitamoto et al. (1990) demonstrate, the error
boxes of the various satellites which may have detected GS 1354$-$64
cover a large area on the sky; if the system were to enter another
`Cen X-2 type' outburst then it would confirm that the source can
indeed display such a wide range of X-ray spectral behaviour -- it may
then be possible to confirm whether or not the three sources are one
and the same. 

\begin{table*}
\begin{center}
\caption{Comparison between the five BHC X-ray transients which have remained in the low/hard state throughout an outburst.}
\label{hard-state}
\begin{tabular}{lcccccc}
\hline
Source&Date&X-ray Spectrum&mHz QPO &Radio Spectrum&Optical Peak&Optical/X-ray\\
&&&During Outburst?&&Preceding X-ray Peak?&Luminosity Ratio\\
\hline
GS 1354$-$64&1997&low/hard&$\nu$ increases&flat $\rightarrow$ inverted &Yes&optically bright\\
GS 2023+338&1989&$^1$low/hard (mostly?)&?&$^2$flat $\rightarrow$ inverted&$^3$Yes&$^4$optically bright\\
GRO J0422+32&1992&$^5$low/hard&$^6$mHz QPO present&$^7$flat $\rightarrow$ inverted &$^8$Possibly$^*$&? (no soft X-ray obs.)\\
GRO J1719$-$24&1993&$^9$low/hard&$^9$$\nu$ increases&$^{10}$flat&?&? (no soft X-ray obs.)\\
XTE J1118+480&2000&$^{11}$low/hard&$^{12,13}$$\nu$ increases&$^{14}$inverted&?&$^{15}$optically bright\\
\hline
\end{tabular}
\end{center}
\begin{tabular}{llll}
1. Tanaka \& Lewin (1995)&2. Han \& Hjellming (1992)&3. Chen, Shrader, Livio (1997)&4. Van Paradijs \& McClintock (1995)\\
5. Van der Hooft et al. (1999)&6. Kouveliotou et al. (1992)&7. Shrader et al. (1994)&8. Callanan et al. 1995)\\
9. Van der Hooft et al. (1996)&10. Hjellming et al. (1993)&11. Hynes et al. (2000)&12. Yamaoka et al. (2000)\\
13. Wood et al. (2000)& 14. Dhawan et al. (2000)&15. Garcia et al. (2000)&\\
\hline
\end{tabular}
\hspace*{-2cm}$^*$ The optical peak {\em appeared} to precede that of the BATSE data but coverage was insufficient to be conclusive\\
\end{table*}
 
\subsection{Other X-ray transients reaching only the hard state}

An increasing number of supposed `soft' X-ray transients have been
shown to remain in the low/hard X-ray state, the X-ray spectra showing
little contribution from a soft disc component -- although note that
in some of these systems the fluxes (both X-ray and optical) were not
at all `low'. A list of hard state X-ray transients can be found in
Table~\ref{hard-state}. In comparing these five objects a number of
trends can be seen, some of which may turn out to be indicative of
low/hard state outbursts of the X-ray transients.

We find that an increase in QPO frequency, a flat (or inverted)
synchrotron radio spectrum and an over-bright optical counterpart
(i.e. a surprisingly high optical:X-ray luminosity ratio; see Van
Paradijs \& McClintock 1995) may be properties common to the low/hard
state X-ray transients -- although not exclusively so, as an increase
in QPO frequency appears to be a common feature of the soft X-ray
transients also (e.g. GS 1124$-$68, Van der Klis 1995) -- however it
should be noted that the QPOs detected in the soft state tend to occur
at higher frequencies. It is rare for the onset of a soft X-ray
transient event to be observed in the optical but at least two of
these low/hard state events have been observed to peak in the optical
before the X-ray, suggesting that something other than reprocessing of
X-rays dominates the optical emission. While these trends are clearly
tentative we note that none of these systems provides an exception to
the `rule' -- either they follow the trends or there have been
insufficient observations to determine either way.

There is a sixth source, Aql X-1, which also shows some (but not all)
of these trends -- it has been omitted from Table~\ref{hard-state} as
it is a neutron star X-ray transient, its X-ray spectral properties
classifying it as an atoll source (Cui et al. 1998, Reig et
al. 2000). However, Aql X-1 has also been observed to peak in the
optical before the soft X-rays (Shahbaz et al. 1998) and also
demonstrates an increase in QPO frequency throughout the outburst (Cui
et al. 1998). It should, however, be pointed out that the QPOs seen in
Aql X-1 are found at kHz frequencies rather than the mHz QPOs found in
the tabulated sources. Consequently it may be incorrect to compare
them, but see also Psaltis, Belloni, Van der Klis (1999). Unlike the
BHC systems there is no optical excess. Although there was a weak
radio counterpart to the outburst (Hjellming et al. 1990) there was no
spectral information -- Aql X-1 is currently one of only a small
number of atoll sources for which radio counterparts have been
detected.

It is also worth comparing these five sources with the low/hard states
of persistent black hole X-ray binaries. While mHz QPOs have been
reported in these systems (e.g. Vikhlinin et al. 1994 and references
within), a variable QPO frequency is not a typical feature of the
low/hard state for these systems and so it may seem surprising that we
see it here. However, although the transients remain in the low/hard
state the change in QPO suggests that the inner disc radius is
decreasing -- it therefore appears that the transition to the soft
state is initiated but not completed. Therefore this increase in QPO
frequency should probably be considered a feature of the {\em
transition}, rather than of the state itself -- this would be
consistent with the inner radius of the disc changing during the
transition but remaining constant once a stable X-ray state is
reached. This theory could be easily tested for the persistent sources
during a transition, although {\it RXTE} observations of the 1996
transition of Cyg X-1 suggest that it may not be just a simple
increase in frequency with flux (Cui et al. 1997) -- further
investigation of this would be useful. The radio behaviour of the
persistent sources is very similar to that of these five transients --
for example, Cyg X-1 (Brocksopp et al. 1999) and GX 339$-$4 (Corbel et
al. 2000) both emit flat spectrum jets whilst in the low/hard state
(see also Fender (2000b) for comparison between the radio properties
of persistent and transient sources in the low/hard state). This jet
is quenched on reaching the soft state -- this is probably also the
case for soft X-ray transients, although the material ejected during
the state transition will continue to produce bright radio emission,
despite being physically decoupled from the accretion process
(e.g. A0620$-$00, Kuulkers et al. 1999).

\section{Discussion}

Our multiwavelength dataset has enabled us to make a number of
suggestions as to what took place during the outburst of GS 1354$-$64
in 1997 November. The optical and radio lightcurves suggest that there
were two maxima during the outburst, the first prior to our
observations and the second at $\sim$JD2450784, depending on
wavelength. There is also the hint of a third optical rise just
following the X-ray peak (JD2450796) -- by this time it is possible
that the disc is sufficiently bright in X-rays for the third optical
brightening to be the result of X-ray reprocessing in the disc
(although this is not confirmed by the hard X-ray energy
spectrum). For future outbursts, simultaneous X-ray and high time
resolution optical observations would be extremely beneficial;
discovery of the mHz QPO (mentioned in Section 3.1) in the optical
would help to determine the efficiency of the reprocessing of soft
X-rays.

It is clear that the optical lightcurve reached a local maximum {\em
before} the X-rays peaked. While common in dwarf novae outbursts, a
preceding optical peak is not generally seen in X-ray transient events
-- quite possibly only due to the fact that it is the X-ray satellites
which tend to discover them. If this is the case then it is unlikely
that reprocessing of X-rays was the dominant source of optical
emission. (However, it is possible if there is some
non-straightforward disc geometry e.g. GRO J1655$-$40 (Hynes et
al. 1998)). Our comparison of `soft' X-ray transients which remain in
the low/hard X-ray state for the duration of the outburst shows that
GS 2023+338 and possibly GRO J0422+32 were also observed to peak in
the optical before the X-ray. While this may be coincidence in timing
of the observations, we consider the possibility that in these hard
state transients the X-ray emission is at a sufficiently low level for
the inwards travelling instability to dominate the optical emission;
in outbursts with a soft X-ray component reprocessing of X-rays would
dominate instead. This is confirmed by the surprisingly high
optical/X-ray luminosity ratios that these low/hard state sources have
-- in a `normal' soft event the relative X-ray luminosity would be
much higher than in the hard state sources.

If the peak of the X-ray lightcurve corresponds to the time at which
the instability reaches the hot inner regions of the disc then a
disc-crossing time of $\sim$10 days is inferred. For this to be true
then the disc and therefore the orbit must be large. The fact that no
orbital period has been determined for this source also suggests that
the orbit may be large (this may also contribute to the possible
over-bright optical emission mentioned in Section 3.2) -- although it
is more likely that there is insufficient optical data for any
periodicities to be determined. If indeed the orbit {\em is} large
then BW Cir would have to be an evolved star filling its Roche Lobe,
suggesting a similar type of system to GRO J1655$-$40.

This inferred disc-crossing time is very long and it is unlikely that
the disc of a system with such a low soft X-ray luminosity and hard
X-ray spectrum would extend to a small inner radius. It is also
unlikely that the instability could trigger a radio ejection before
reaching the innermost regions of the disc and causing the X-ray
peak. The increase in QPO frequency suggests that the increase in soft
X-ray luminosity is the result of the inner disc radius decreasing;
the surface area of the disc and the temperature of the inner regions
are therefore increasing. This would be consistent with other X-ray
binaries -- the low/hard state is characterised by a large inner disc
radius and this decreases on transition to the high/soft state
(e.g. Esin et al., 1998).

Therefore it appears that the instability moves inwards, emitting in
the optical regime. As the transported mass fills the `hole' at the
centre of the disc on a viscous timescale, the soft X-ray emission
from the corona increases. This scenario is comparable with that
suggested by Shahbaz et al. (1998) to explain the optical/X-ray delay
of the 1997 outburst of the neutron star X-ray transient Aql X-1.

If we assume that a radio event is triggered when the instability
reaches the inner edge of the disc then this considerably reduces the
required disc crossing time (i.e. the disc crossing time becomes the
delay between the optical and radio peaks rather than the delay
between the optical and X-ray) and seems more likely. From
observations of other sources in the hard state it is probable that
the first radio event is a discrete ejection, with relativistic bulk
velocity, which rapidly evolves to an optically thin
spectrum. Subsequently, in the low/hard state, a powerful, steady,
partially self-absorbed jet is produced which is observed
approximately as a flat-spectrum component (Fender 2000b). It is also
possible that repeated discrete ejections take place -- however, these
are more commonly seen in the soft X-ray transient outbursts and do
not seem to be a common feature of the low/hard X-ray state.

Alternatively (or additionally?), we also note that the optical and
radio data have comparable fluxes. It is therefore possible that an
ejection event took place, emitting synchrotron radiation at optical
and increasingly longer wavelengths as the ejected material became
optically thin at increasing distances from the central source. The
flat spectrum (Fig.~\ref{spectrum}), apparently extending from radio
wavelengths to the $R$ band, would certainly support this
scenario. If, as our value of $\xi$ suggests, the optical component is
over-bright then synchrotron should not be discounted as a possible
contributor to the optical emission in addition to the thermal disc
spectrum.

It is important to consider the implications of a jet (or mass
ejection) that emits flat spectrum synchrotron radiation over such a
wide range of frequencies. While there are a number of models that
work reasonably well for the `flat spectrum' AGN, none of these is
consistent with the {\em much flatter} radio -- (sub-)mm spectra of
e.g. Cyg X-1, GRS 1915+105 and Cyg X-3 (Fender et al. 2000 and
references within). The most convincing explanation for the flat
spectrum is synchrotron radiation from a partially self-absorbed
jet. With the power of the jet directly proportional to the bandwidth,
detection of a high frequency cut-off to the flat spectrum would be
very useful in determining the power of the jet -- this has not been
found in any of the flat spectrum X-ray binaries.  

In summary, our limited observations suggest that the 1997 outburst of
GS 1354$-$644 was the result of an instability crossing the disc on a
slow viscous timescale, emitting at optical wavelengths. As the
instability reached the (large) inner radius of the disc it triggered
a mass ejection which emitted synchrotron radiation at radio
wavelengths and possibly through to the near infrared (and optical?)
regime. The X-ray source remained in the low/hard state throughout,
gradually becoming brighter as the instabilities carried more matter
into the hot inner regions and subsequently decreasing the inner disc
radius. These results clearly reflect the importance of simultaneous
optical, infrared and radio studies of future X-ray transient events.

\section*{acknowledgements}

This work was completed on the Sussex {\sc STARLINK} node. We are very
grateful for the quick-look results provided by the {\it ASM}/RXTE and
{\it CGRO}/BATSE teams and to the various people who have taken
observations for us. The Australia Telescope is funded by the
Commonwealth of Australia for operation as a National Facility managed
by CSIRO. We are also grateful to Ben Stappers, J\"orn Wilms and
Guillaume Dubus for useful conversations.

CB acknowledges a PPARC studentship; PGJ and PJG are supported by NWO
Spinoza grant 08-0 to E.P.J. Van den Heuvel. PJG is also supported by
a CfA Fellowship.


\begin{thebibliography}{99}
\bibitem[belloni]{belloni}Belloni T., Hasinger G., 1990, A\&A, 227, L33
\bibitem[Brocksopp 1999]{brocksopp} Brocksopp C., Fender R.P., Lariononv V., Lyuty V.M., Tarasov A.E., Pooley G.G., Paciesas W.S., Roche P., 1999, MNRAS, 309, 1063
\bibitem[Buxton et al.]{buxton}Buxton M., Vennes S., Ferrario L., Wickramasinghe D.T., 1998, IAUC 6815
\bibitem[callanan]{callanan}Callanan et al., 1995, ApJ, 441, 786
\bibitem[cardelli]{cardelli}Cardelli J.A., Clayton G.C., Mathis J.S., 1989, ApJ, 345, 245  
\bibitem[Castro-Tirado 1997]{castro} Castro-Tirado A.J., Ilovaisky S., Pederson H., 1997, IAUC 6775
\bibitem[Charles 1998]{charles}Charles P.A., 1998, ``Theory of Black Hole Accretion Disks'', Eds. M. Abramowicz, G. Bjornsson \& J. Pringle, CUP
\bibitem[chen]{chen} Chen W., Shrader C.R., Livio M., 1997, ApJ, 491, 312
\bibitem[corbel]{corbel} Corbel S., Fender R.P., Tzioumis A.K., Nowak M., McIntyre V., Durouchoux P., Sood R., 2000, A\&A, 359, 251
\bibitem[Cui et al. 1998]{cui98} Cui W., Barret D., Zhang S.N., Chen W., Boirin L., Swank J., 1998, ApJ, 502, L49
\bibitem[Cui et al. 1999]{cui99} Cui W., Zhang S.N., Chen W., Morgan E.H. 1999, ApJ, 512, L43
\bibitem[cui]{cui} Cui W., Zhang S.N., Focke W., Swank J.H., 1997, 484, 383
\bibitem[dhawan]{dhawan}Dhawan V., Pooley G.G., Ogley R.N., Mirabel I.F., 2000, IAUC 7395
\bibitem[Ebisawa et al. (1994)]{ebisawa} Ebisawa K., et al., 1994, PASJ, 46, 375
\bibitem[esin]{esin}Esin A.A., Narayan R., Cui W., Grove J.E., Zhang S.N., 1998, ApJ, 505, 854
\bibitem[Fender 2000]{fender2}Fender R.P., 2000a, To be published in Proc. ESO workshop `Black Holes in binaries and galactic nuclei', Eds Kaper L., Van den Heuvel E.P.J. and Woudt P.A., Springer-Verlag
\bibitem[Fender 2000]{fender2}Fender R.P., 2000b, MNRAS in press (astro-ph/0008447)
\bibitem[Fender et al. 2000]{fender} Fender R.P., Pooley G.G., Durouchoux P., Tilanus R.P.J., Brocksopp C., 2000, MNRAS, 312, 853
\bibitem[Fender et al. 1997]{fender} Fender R.P., Tingay S,J., Higdon J., Wark R., Wieringa M., 1997, IAUC 6779
\bibitem[Fender et al. 1999]{fender} Fender et al., 1999, ApJ, 519, L165
\bibitem[fishman et al. 1989]{fishman} Fishman G.J. et al., 1989, Proc. GRO Science Workshop, NASA, Ed. Johnson W.N.
\bibitem[Francey 1971]{francey} Francey R.J. 1971, nature Phys. Sci, 229, 228
\bibitem[FKR 1992]{fkr}  Frank J., King A., Raine D., 1992, `Accretion Power in Astrophysics', Chap. 5
\bibitem[garcia]{garcia}Garcia M., Brown W., Pahre M., McClintock J., Callanan P., Garnavich P., 2000, IAUC 7392 
\bibitem[han]{han}Han X., Hjellming R.M., 1992, ApJ, 400, 304
\bibitem[Harmon \& Robinson]{harmon} Harmon B.A., Robinson C.R., 1997, IAUC 6774
\bibitem[hjellming]{hjellming}Hjellming R.M., Han X., Roussel-Dupre, 1990, IAUC 5112
\bibitem[hjellming]{hjellming}Hjellming R.M., Rupen M.P., Shrader C.R., Campbell-Wilson D., Hunstead R.W., McKay D.J., 1993, ApJ, 470, L105
\bibitem[hjellming]{hjellming}Hjellming R.M. et al., 1999, ApJ, 514, 383
\bibitem[homan]{homan}Homan J., Wijnands R., Van der Klis M., Belloni T., Van Paradijs J., Klein-Woldt M., Fender R.P., M\'endez M., 2000, ApJ submitted
\bibitem[Hynes]{hynes}Hynes R.I., 1998, New Astronomy Reviews, 42, 605
\bibitem[Hynes]{hynes98}Hynes R.I., O'Brien K.O., Horne K., Chen W., Haswell C.A., 1998, MNRAS, 299, L37
\bibitem[Hynes]{hynes98}Hynes R.I., Mauche C.W., Haswell C.A., Shrader C.R., Cui W., Chaty S., 2000, ApJL submitted
\bibitem[Jahoda et al. 1996]{jahoda} Jahoda K., Swank J.H., Giles A.B., Stark M.J., Strohmayer T., Zhang W., Morgan E.H. 1996, SPIE, 2808, 59
\bibitem[Jonker et al. 2000]{jonker} Jonker P.G., Van der Klis M., Wijnands R., Homan J., Van Paradijs J., M\'endez M., Ford E.C., Kuulkers E.,\& Lamb F.K. 2000, ApJ, in press
\bibitem[Kitamoto et al. (1990)]{kitamoto} Kitamoto S., Tsunemi H., Pedersen H., Ilovaisky S.A., Van der Klis M., 1990, ApJ, 361, 590
\bibitem[Kouveliotou]{Kouveliotou}Kouveliotou et al., 1992, IAUC 5592
\bibitem[Kuulkers et al. 1999]{kuulkers}Kuulkers E., Fender R.P., Spencer R.E., Davis R.J., Morison I, 1999, MNRAS, 306, 919
\bibitem[Landolt 1992]{landolt}Landolt A.U., 1992, ApJ, 140, 340 
\bibitem[levine] {levine} Levine A.M., Bradt H., Cui Wei., Jernigan J.G., Morgan E.H., Remillard R., Shirey R.E., Smith D.A., 1996, ApJ., 469, L33
\bibitem[Makino et al. 1987]{makino} Makino F. et al., 1987, IAUC 4342
\bibitem[Markert et al. 1977]{markert} Markert T.H., et al. 1979, Ap. J. Suppl., 39, 573
\bibitem[M\'endez, Belloni and Van der Klis (1998)]{mariano} M\'endez M., Belloni T., Van der Klis M., 1998, ApJ, 499, 187
\bibitem[orosz]{orosz}Orosz J., 1997, ApJ, 477, 876
\bibitem[Psaltis, Belloni, \& Van der Klis 1999]{psaltis} Psaltis D., Belloni T., Van der Klis M. 1999, ApJ, 520, 262
\bibitem[reig]{reig}Reig P., M\'endez M., Van der Klis M., Ford E.C., 2000, ApJ, 530, 916
\bibitem[Remillard, Marshall \& Takeshima (1997)]{rmt} Remillard R., Marshall R., Takeshima T., 1997, IAUC 6772
\bibitem[Revnivtsev et al. (2000)]{revnivtsev} Revnivtsev M.G., Borozdin K., Priedhorsky W.C, Vikhlinin A., 2000,ApJ, 531
\bibitem[shahbaz]{shabaz} Shahbaz T., Bandyopadhyay R.M, Charles P.A., Wagner R.M., Muhli P., Hakala P., Casares J.,Greenhill, J., 1998, MNRAS, 300, 382
\bibitem[shrader]{shrader} Shrader C.R., Wagner R.M., Hjellming R.M., Han X.H., Starrfield S.G., 1994, ApJ, 434, 698
\bibitem[soria et al.]{soria} Soria R., Bessell M.S., Wood P., 1997, IAUC 6732
\bibitem[sunyaev]{sunyaev}Sunyaev et al. 1993, A\&A, 280, L1 
\bibitem[Swinbanks 1987]{swinbanks}Swinbanks D., 1987, Nature, 326, 322
\bibitem[Tanaka and Lewin (1995)]{tan} Tanaka Y, Lewin W.H.G., 1995, `X-Ray Binaries' Chap 3, Eds. Lewin W.H.G., Van Paradijs J., Van den Heuvel E.
\bibitem[Tanaka and Shibazaki (1996)]{tan2} Tanaka Y, Shibazaki N., 1996, ARAA, 34, 607
\bibitem[Van der Hooft et al. (1996)]{frank} Van der Hooft F., et al. 1996, ApJ, 458, L75
\bibitem[Van der Hooft et al. (1999)]{frank99} Van der Hooft F., et al. 1999, ApJ, 513, 477
\bibitem[Van der Klis (1994)]{mich94} Van der Klis M., 1994, A\&A, 283, 469
\bibitem[Van der Klis (1995)]{mich95} Van der Klis M., 1995, `X-Ray Binaries' Chap 6, Eds. Lewin W.H.G., Van Paradijs J., Van den Heuvel E.
\bibitem[Van Paradijs 1996]{jvp1} Van Paradijs J., 1996, ApJ, 464, L138
\bibitem[Van Paradijs 1995]{jvp+mcclintock} Van Paradijs J., McClintock J.E., 1995, `X-Ray Binaries' Chap 2, Eds. Lewin W.H.G., Van Paradijs J., Van den Heuvel E.
\bibitem[Van Paradijs 1998]{jvp} Van Paradijs J., 1998, `The Many Faces of Neutron Stars', page 279, Eds. Buccheri R., Van Paradijs J., Alpar M.A.
\bibitem[Vikhlinin]{vikhlinin} Vikhlinin A. et al., 1994, ApJ, 424, 395
\bibitem[Wijnands \& Van der Klis 1999]{wijnands1999} Wijnands R., \& Van der Klis M. 1999, ApJ, 514, 939
\bibitem[Wood]{wood} Wood K.S. et al., 2000, ApJL submitted
\bibitem[Yamaoka]{Yamaoka}Yamaoka K., Ueda Y., Dotani T., Durouchoux P., Rodriguez J., 2000., IAUC 7427
\bibitem[zycki]{zycki}\.Zycki P.T., Done C., Smith D.A., 1999, MNRAS, 309, 561

\end{thebibliography}
\end{document}